\documentclass{aipproc}

\layoutstyle{6x9}

\begin{document}

\title 
{
	Exploring the Charm Sector with CLEO-c
}

\author{D. Urner}{
	address={Cornell University, Wilson Lab, Ithaca NY 14853, USA }}
	
\begin{abstract}
The CLEO collaboration proposes to explore the charm sector starting early 2003. 
It is foreseen to collect on the order of 6 million $D\overline{D}$ pairs, 300000 $D_s\overline{D}_s$ 
pairs at threshold and one billion $J/\psi$ decays. High precision charm data will enable us to 
validate upcoming Lattice QCD calculations that are expected to produce 1-3\% errors for some 
non-perturbative QCD quantities. These can then be used to improve the accuracy 
of CKM elements. The radiative $J/\psi$ decays will be the first high statistics 
data set well suited for meson spectroscopy between 1600 and 3000 MeV.
\end{abstract}

\maketitle

\section{Introduction}
Let's start with a lofty goal: We strive for the mastery of a non-perturbative strongly 
coupled theory: QCD. Couplings in field theory do not typically have to be weak. Indeed, strong 
interactions are the expected phenomena if one reaches beyond the Standard Model. 
It will therefore be of great benefit if we can understand the effects of strong couplings 
in QCD. High precision predictions of QCD will also remove road blocks for many weak and flavor
physics measurements. 

Lattice QCD has matured over the last decade. We finally can expect the first non-perturbative QCD 
results with 1-3\% errors. CLEO-c will provide crucial data in a timely fashion to validate them
and help guide the theory on its long way from easier predictions to a full understanding of 
non-perturbative QCD effects. This will result, for example, in improved measurements of $V_{cs}$ 
and $V_{cd}$ at the 1\% level. CLEO-c data will also provide a large number of basic measurements 
needed in heavy flavor physics and future efforts in understanding physics beyond the standard 
model. A detailed description can be found in~\cite{citcleoc}.

\subsection{Data Sets}
We plan to acquire the CLEO-c data in a 3 year program. The use of present 
and future CLEO data sets will be considered in the course of this paper. The expected size of 
the data sets are shown in Table~\ref{tabledatasets}. 

CLEO-c intends to accumulate 30 x 10$^6$ events at $\psi ''$ and about 1.5 x 10$^6$ events at 
$\psi$(4140). The number of expected $D_s\overline{D}_s$ events is uncertain within a factor of two
because of conflicting earlier measurements. This will be clarified with an early scan, which 
will determine the point of largest $D_s\overline{D}_s$ production. We expect to collect a total 
of about one billion $J/\psi$ events. Smaller data sets are considered at the $\tau\tau$ 
threshold (3557 MeV) at the $\psi'$(3686) at $\Lambda_c \overline{\Lambda}_c$ threshold (5200 MeV) 
and a scan over the full 3-7 GeV region.

In 2002, before lowering the energy to the charm sector, CLEO plans to collect data at the three
narrow 1S, 2S and 3S $\Upsilon$ resonances with an integrated luminosity of about 1fb$^{-1}$ each.

The CLEOII and CLEOIII data sets gathered at the $\Upsilon$(4S) contain a large number 
of 2-photon events.

\begin{table}
\label{tabledatasets}
\begin{tabular}{lcc}\hline
Center of mass energy & Luminosity & Decays/Physics \\ \hline
$\Upsilon$(4S)(10580) & 24 fb$^{-1}$ & 2-photon physics \\ \hline
$\Upsilon$(3S)(10355) & 1 fb$^{-1}$ & $\eta_b$ \\
$\Upsilon$(1S)(9460) & 1 fb$^{-1}$ & meson spectroscopy \\
$\Upsilon$(2S)(10023) & 1 fb$^{-1}$ &  \\ \hline
$\psi''$(3770) & 3 fb$^{-1}$ & 6 x 10$^6$ $D\overline{D}$ \\
$\psi$(4140) & 3 fb$^{-1}$ & 3.0 x 10$^5$ $D_s\overline{D}_s$ \\
$J/\psi$(3097) & 1 fb$^{-1}$ & 6.0 x 10$^7$ radiative $J/\psi$ decays  \\ \hline

\end{tabular}
\caption{Size of datasets considered in the discussion of this paper. The data at the $\Upsilon$(4S) 
represents the CLEOII and CLEOIII data sets. The data sets at the narrow $\Upsilon$ resonances 
are taken just prior to CLEO-c.}
\end{table}

\subsection{Accelerator: Modifications to CESR}
\begin{figure}
\resizebox{.3\textwidth}{!}{\includegraphics{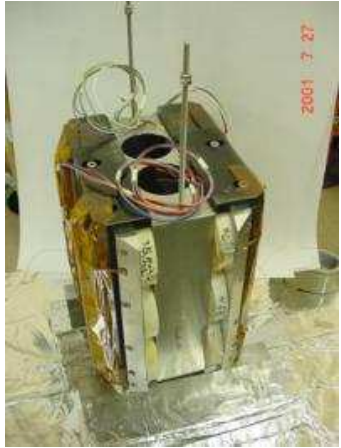}}
\label{figwiggler}
\caption{A 3 pole test module for the super conducting wigglers needed in the upgrade to CLEO-c}.
\end{figure}

For the upgrade to CLEO III new superconducting quadrupoles for the final focusing system were 
built. They prove to be crucial, since they enable us to lower the beam energy and run in the 
region of the charm system. Accelerators find that the luminosity typically scales at best with 
L$\sim$E$_b^4$. This behavior can be changed if one introduces wigglers, which will cool the beam 
transversely to ideally a linear correlation of luminosity and beam energy. We plan to build 14 
superconducting wiggler modules, each having 1.3 m of length, a peak field of 2T, and a 40 cm period. 
A 3-pole test module is shown in figure~\ref{figwiggler}. The projected beam spread at 
the J/$\psi$ will be about: $\Delta$E$_b \sim$1.2 MeV. The expected machine performance is shown in 
Table~\ref{tabcesrperformance}.

\begin{table}[b]
\label{tabcesrperformance}
\begin{tabular}{lc} \hline
$\sqrt{s}$ & L(10$^{32}$cm$^{-2}$s$^{-1}$)\\ \hline
4.1 GeV & 3.6 \\
3.77 GeV & 3.0 \\
3.1 GeV	& 2.0\\
\end{tabular}
\caption{Expected machine performance for CLEO-c.}
\end{table}

\subsection{Detector: CLEO III becomes CLEO-c}
The CLEO III detector is a wonderful detector to study the charm system. The tracking system and 
the calorimeter cover 93\% of the solid angle, while the ring-imaging \u{C}erenkov counter (RICH) covers 83\% of the solid angle. 

The tracking system consists of the main drift chamber~\cite{citdr} for which we find a hit resolution 
of 88 $\mu$m. The 4 layer silicon detector~\cite{citsil} has prematurely degraded, which is observed as a dramatic
efficiency loss in the sensors signals. The origin of this problem is yet unknown.  Cornell is building a 6-layer 
high angle stereo drift chamber to replace the silicon detector. For low momentum tracks, such as those typically generated 
when running at J/$\psi$ energies, the performance of this drift chamber is comparable to a silicon vertex detector, 
because multiple scattering is the dominant contribution to the track resolution. We plan to run with a reduced B field of 1T 
(1.5 Tesla for $\Upsilon$ running). That leads to a resolution of 0.35\% at 1 GeV for charged tracks. 

The calorimeter~\cite{citcalo} consists of 7800 Cesium Iodide crystals and measures the photons with a 
$\frac{\sigma_E}{E}$ = 2\% at 1 GeV and 4\% at 100 MeV. 

Particle identification can be done with dE/dx with a resolution of 5.7\% for 
minimum ionizing pions. A ring imaging \u{C}erenkov counter~\cite{citrich} has been installed with the CLEOIII upgrade. 
It has been shown to perform excellently. It covers 83\% of the solid angle. For 0.9 GeV particles
the kaons are identified with 87\% efficiency and a pion fake rate of 0.2\%. 

Data from the drift chamber and the calorimeter are used in the trigger. It is pipelined with a 
latency of 2.5 $\mu$s. The trigger is fully programmable and can easily be adapted to the
new event signatures.

The data acquisition system can accept hardware triggers up to 1 kHz and is designed to write data 
to tape with a speed of about 300 Hz. The event size is 25 kB and the data throughput is 6Mb. 
This means that the data acquisition infrastructure will be able to handle the large data rates at 
the J/$\psi$ peak.

\section{Running at Threshold}
\begin{figure}
\label{figdtags}
\resizebox{.8\textwidth}{!}{\includegraphics{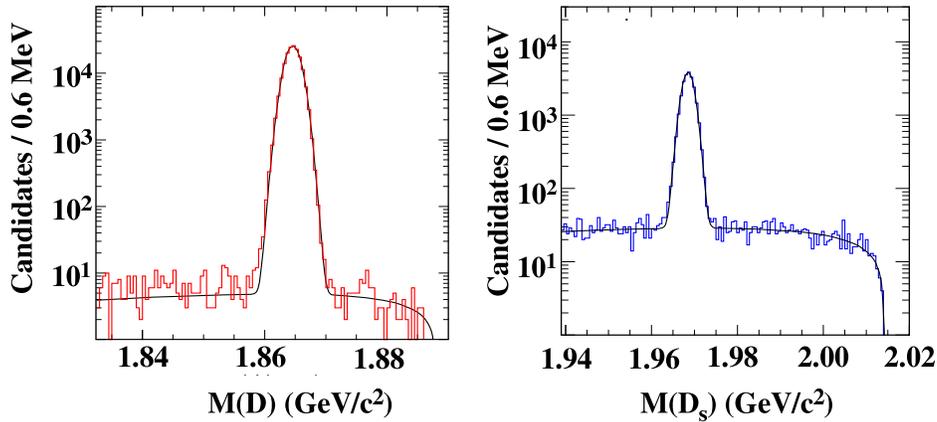}}
\caption{MC events equivalent to 1fb$^{-1}$ of data. Left: D$\rightarrow$K$\pi$ tags. The width 
of the D peak is 1.3 MeV/c$^2$. Right: D$_s\rightarrow$KK$\pi$ tags. The width of the D$_s$ peak 
is 1.4 MeV/c$^2$.}
\end{figure}

There are important advantages to running at the open flavor thresholds.
Large cross sections mean the data can be acquired in one run period rather than over many years. 
The multiplicity of particles in the final state is smaller, which reduces backgrounds and 
increases efficiencies. We expect very clean data samples. Single tag events can be used to 
constrain $\nu$ reconstruction, double tags will be used for hadronic measurements. We 
anticipate 6 million D tags and 300,000 D$_s$ tags. The signal-to-background ratio for the 
D$\rightarrow$K$\pi$ tag is estimated to be S/B: 5000 and for D$_s\rightarrow$KK$\pi$ tag: 
S/B $\sim$ 100, (see Figure~\ref{figdtags}). Many of the analyses can profit from using kinematic 
constraints at threshold. This can, for example, lead to a reduced systematic error if one 
can forego lepton identification. Further advantages are that the initial states 
are pure (no fragmentation) and that the D's are coherently produced.

\subsection{Absolute Hadronic Charmed Hadron Branching Fractions}
Absolute branching fractions are important since, for a lot of analyses at the highest energies as well as 
in the $B$-system, an inaccurate knowledge of D, D$_s$,... decays can result in large systematic errors. 
Using double-tagged events at threshold leaves only major systematic error contributions from efficiency
uncertainties in the tracks and showers. An overview of the expected results is shown in Table~\ref{tababsbranch}.

\begin{table}
\label{tababsbranch}
\begin{tabular}{lccccc} \hline
Particle & \# of double tags & Statistical & Systematic & Background & Total   \\ 
         &                  & Error       & Error      & Error      & Error   \\ \hline
D$^0$    & 53,000           & 0.4\%       & 0.4\%      & 0.06\%     & 0.6\%\\
D$^+$    & 60,000           & 0.4\%       & 0.6\%      & 0.10\%     & 0.7\%\\
D$_s^+$  & 6,000            & 1.3\%       & 1.1\%      & 0.90\%     & 1.9\%\\
$\Lambda_c$& 17,000         & $\sim$4\%       & small      & small      & $\sim$4\%\\

\end{tabular}
\caption{Total number of double tag events and precision on absolute charm branching fractions, 
assuming 3fb$^{-1}$ data at $D\overline{D}$, $D_s\overline{D}_s$ thresholds and 1fb$^{-1}$ 
at $\Lambda_c \overline{\Lambda}_c$ threshold}
\end{table}

In the case of D decays the statistics are high enough that we can concentrate on the most simple D decays. 
For one fb$^{-1}$ we expect 1500 D$^0\rightarrow$K$^+\pi^-$ events with no background and 
8446 D$^+\rightarrow$K$^-\pi^+\pi^+$ with 25 background events. 
For D$_s$, the decays into K$^-$K$^+\pi^-$, K$^-$K$^+\pi^-\pi^0$, $\eta\pi^-$, 
$\eta\rho^-$, and $\eta'\pi^-$ were considered and combinations with less than 20\% background were used to 
generate the numbers shown in Table~\ref{tababsbranch}.

The charmed baryon resonances measured with collisions at a large center of mass are typically presented as branching ratios 
with respect to the $\Lambda_c^+\rightarrow$pK$^-\pi^+$ decay. With one fb$^{-1}$ at the threshold (4.6 GeV), one expects 
about 500 $\Lambda_c\overline{\Lambda}_c$ double-tagged events, a rough estimate since neither the cross section 
nor the $\Lambda_c^+\rightarrow$pK$^-\pi^+$ branching fraction are well known. 

\subsection{Meson Decay Constant}
The hadronic physics for leptonic decays of the D$_q$ and B$_q$ mesons is encapsulated in single non-perturbative QCD parameters f$_q$. 
Given the knowledge of their values, one can extract |V$_{cs}$|, |V$_{cd}$|, |V$_{ts}$| and |V$_{td}$|. Our current knowledge of the 
uncertainty in f$_{D_s}$ and f$_D$ is 35\% and 100\%, respectively, while f$_B$ and f$_{B_s}$ will not be measured in the 
foreseeable future. Lattice QCD should, however, be able to calculate the ratio of $\frac{f_B}{f_D}$ very accurately, so that the 
determination of f$_D$ will indirectly help in the extraction of |V$_{ts}$| and |V$_{td}$|.

The leptonic D and D$_s$ decay branching fractions are sizable and enable a direct determination of the charm meson 
decay constant from the measurement of D$^+\rightarrow\mu^+\nu$, D$_s^+\rightarrow\mu^+\nu$, and D$_s^+\rightarrow\tau^+\nu$.
If one assumes unitarity of the CKM matrix to constrain the values of |V$_{cd}$| and |V$_{cs}$| one can measure
$\frac{\delta f_{D_s}}{f_{D_s}}$ = 2.1\% and $\frac{\delta f_{D}}{f_{D}}$ = 2.6\%.

\subsection{Semileptonic Form Factors and Determination of |V$_{cs}$| and |V$_{cd}|$}
\label{secvcs}
The semileptonic from factors $|f_+(q^2)|^2$ encapsulate the hadronic physics of semileptonic decays
\begin{equation}
\frac{d\Gamma}{dq^3} = \frac{G_F^2}{24\pi^3}|V_{cs}|^2 p_K^3 |f_+(q^2)|^2
\end{equation}
in the example of a $c \rightarrow s$ transition. If we again assume 3 generation unitarity one can extract the 
semileptonic form factors from processes like D$^0\rightarrow\pi$e$\nu$ or D$^+\rightarrow$K$^{*0}$e$^+\nu$. 
The estimated precision for the parameters of the pseudoscalar to pseudoscalar and pseudoscalar to vector form factors are shown in 
Table ~\ref{tabsemformfact}. 

\begin{table}
\label{tabsemformfact}
\begin{tabular}{lcccccc}
Decay                                     & PDG2000          & CLEO-c           & Form Factor         & Form Factor               & expected  & CKM       \\ 
Mode                                      & ($\delta$B/B \%) & ($\delta$B/B \%) & Type                & Parameter                 & Precision & Element   \\ \hline
D$^0\rightarrow$K$^-$e$^+\nu$             & 5                & 0.4              &                     &                           &           & |$V_{cs}$| \\
D$_s^+\rightarrow\phi^-$e$^+\nu$          & 25               & 3.1              &                     &                           &           & |$V_{cs}$| \\
D$^+\rightarrow\pi^0$e$^+\nu$             & 48               & 2.0              &                     &                           &           & |$V_{cd}$| \\
D$^0\rightarrow\pi^-$e$^+\nu$             & 16               & 1.0              & PS $\rightarrow$ PS & $\delta$f$_+$(0)/f$_+$(0) & $\sim$1\%     & |$V_{cd}$| \\
                                          &                  &                  &                     & slope                     & $\sim$4\%     &           \\
D$^+\rightarrow\overline{K}^{*0}$e$^+\nu$ & 9                & 0.6              & PS $\rightarrow$ V  & $\delta$A$_1$(0)/A$_1$(0) & $\sim$2\%     & |$V_{cs}$| \\
                                          &                  &                  &                     & $\delta$A$_2$(0)/A$_2$(0) & $\sim$5\%     &           \\
                                          &                  &                  &                     & $\delta$V(0)/V(0)         & $\sim$5\%     &           \\

\end{tabular}
\caption{This table contains uncertainties on the branching fractions for several D and D$_s$ decay modes and precision of 
semileptonic form factor parameters.}
\end{table}

|V$_{cd}$| is related to the decay rate $\Gamma$, which we can calculate from the absolute branching ratio 
$B($D$^0\rightarrow$K$^-$e$^+\nu)$  and the mean lifetime $\tau_{D^0}$ as:
\begin{equation}
	\Gamma(D^0\rightarrow K^-e^+\nu) = \frac{B(D^0\rightarrow K^- e^+ \nu)}{\tau_{D^0}} = T_d|V_{cs}|^2
\end{equation}
T$_d$ is taken from theory and requires the knowledge of the PS to PS semileptonic form factor. We can expect lattice QCD 
calculations with an error of $\delta$T$_s$/T$_s$ = 3\% within a few years as discussed further below. This will result in a 
precision for |V$_{cd}$| of 1.7\%. In a similar way, |V$_{cs}$| can be determined with a precision of 1.6\%. Both CKM matrix elements
can also be extracted from leptonic decays. Combining leptonic and semileptonic measurements we expect final precisions of
1.4\% and 1.1\% on |V$_{cd}$| and |V$_{cs}$|, respectively.

\section{Mapping out the $\Upsilon$ and $J/ \psi$ systems}

Starting late fall 2001 CLEO will collect about 1 fb$^{-1}$ on each of the $\Upsilon$(1S) - $\Upsilon$(3S) resonances. 
Most present theories~\cite{citdiscoveretab} 
indicate that the ground state of the $\Upsilon$ system, the $\eta_b$, could be discovered via the hindered M1 transition 
from the $\Upsilon$(3S) state. One also would expect to see the decay $\Upsilon$(3S)$\rightarrow \pi^+\pi^-$h$_b$~\cite{cithb}. 
If found, its large predicted decay branching fraction into $\eta_b$ of 50\% opens a further avenue to observe the $\eta_b$.

CLEO also should be able to observe the 1$^3$D$_J$ states. The $b\overline{b}$ system is unique in that it has states with L=2 that 
lie below the open-flavor threshold. We expect unquenched lattice calculations soon for the center of gravity of the triplets of
both D states to about $\sim$5 MeV. Current theoretical calculations and our existing data suggest that we can expect 20 - 40 fully 
reconstructed events in the decay $\Upsilon$(3S)$\rightarrow \gamma_1\chi'_b\rightarrow\gamma_1\gamma_2(^3D_J)$ $\rightarrow
\gamma_1\gamma_2\gamma_3\chi_b\rightarrow\gamma_1\gamma_2\gamma_3\gamma_4\Upsilon(1S)
\rightarrow\gamma_1\gamma_2\gamma_3\gamma_4 l^+ l^-$, which should enable the extraction of the center of gravity of the triplet of
the 1$^3$D$_J$ state to about 3 MeV. From this, the mass of the lowest state can be predicted and a scan can directly establish its 
mass, gaining a measure of the S-D mixing in the $b\overline{b}$ system.

In the J/$\psi$ system only very few states are measured with high precision. Although we know the ground state $\eta_c$, its width 
is measured very poorly. The $\eta_c'$ and h$_c$ states need confirmation. The $^{3,1}$D$_J$ and 2$^{3,1}$P$_J$ states still have to
be found. The region above the $D\overline{D}$ threshold is generally explored very little, despite the fact that a lot of interesting 
physics might be extracted if one could identify, for example, charmed hybrid states. CLEO-c will have to make a scan in order to find 
the energy with the largest $D_s\overline{D}_s$ decay rate. A more detailed scan between 3.6 GeV and 5 GeV is also considered.

\section{The Bright Future of Lattice QCD}
\label{seclattice}
Lattice QCD is a full implementation of QCD and can therefore in principle produce accurate results, also for the low energy phenomena
that cannot be treated perturbatively. It has, however, failed to make good predictions with well understood errors because
of technical difficulties~\cite{citlat1}. In the last few years, some real breakthroughs have been achieved~\cite{citlat2}, so that we can expect some of the
theoretically easier calculations to appear in the next few years with sound error estimates of 1-2\%. 

In general, one was able to incorporate known features of QCD into the lattice calculations. Perturbation theory is used to describe 
short distance physics and connect the lattice to the continuum. Second order results lead to relative errors of O($\alpha_s^3$).
One would like to keep lattice spacing as large as possible in order to minimize computer time. Improved discretizations remove
errors by adding correction terms. These and other improvements are required in order that one can finally get unquenched results, which 
include effects from $q\overline{q}$ loops. When this can be done with realistic d,u-quark masses, the errors on the results become
quantifiable and errors of order 2\% are achievable for many calculations. Examples are masses, decay constants, semileptonic form 
factors, and mixing amplitudes for D, D$_s$, D$^*$, D$_s^*$, B, B$_s$, B$^*$, B$_s^*$, and corresponding baryons; masses, 
leptonic widths, electromagnetic form factors, and mixing amplitudes for any meson in $\psi$ and $\Upsilon$ families below the 
open flavor threshold. 

It should be stressed that lattice QCD can make all these predictions using only the quark masses and $\alpha_s$ as input 
parameters. Today there are not enough measurements in the one percent region for the quantities mentioned above. CLEO-c, however, will 
be able to provide most of them accurately enough, enabling us to validate the new lattice results and methods. Particularly the
measurements in the $\psi$ and $\Upsilon$ region provide an excellent test ground. This validation 
is needed so that lattice QCD results can be trusted and used to increase the accuracy of theoretical predictions for many 
different aspects of physics.

\subsection{Impact of CLEO-c Results and Lattice QCD on Flavor Physics}

An immediate application of lattice QCD results by CLEO-c involves the extraction of  |V$_{cs}$| and |V$_{cd}$| 
by the use of the meson decay form factor and the measurement of leptonic decay branching fractions of D and D$_s$,
as well as the semileptonic form factors and the measurement of the semileptonic decay branching fractions of D and D$_s$, 
described above.

Even before a full validation of lattice QCD, measurements of the D-meson form factor will result in better predictions of the 
B-meson form factor, which will tighten the constraints on the unitarity triangle from $B\overline{B}$ and $B_s\overline{B}_s$ 
mixing measurements. Using SU(3) and heavy quark symmetry the semileptonic form factor for the process B$\rightarrow\pi$e$^+\nu$ 
can be predicted from the measurement of the semileptonic form factor extracted from the process D$^+\rightarrow$K$^{*0}$e$^+\nu$, 
which will result in an improved |V$_{ub}$| determination.

\begin{figure}
\label{figunitaritytriangle}
\resizebox{.5\textwidth}{!}{\includegraphics{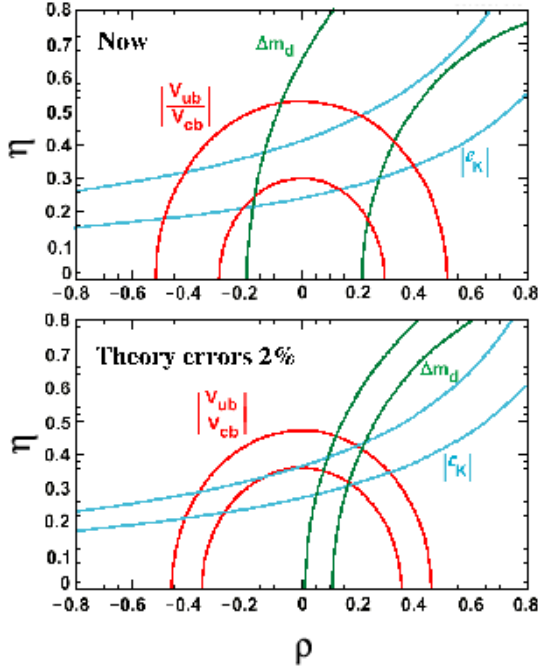}}
\caption{Above: Current situation taken from~\cite{cithock}. Below: Prediction on the limits using 2\% errors on form factors and today's data.}
\end{figure}

Finally, the impact of 2\% theoretical errors on form factors produced by a validated lattice QCD theory are shown 
in Figure~\ref{figunitaritytriangle} using today's measurements.
It is obvious that only an increased bound from |V$_{ub}$|  and $B\overline{B}$ mixing results together with the sin(2$\beta$) 
measurements will be able to give a significant answer on the question if the unitarity triangle is indeed closed or 
if the effect of new physics is observed. The limits from the $\epsilon_K$ measurement will also be improved considerably. 

\section{Meson Spectroscopy from $J/ \psi$ decays}
Lattice QCD will have answers for some questions in the near future, however there are many quantities that will remain difficult
to calculate for some time. There are only some ideas on how to treat effects like mixing, or the inclusion of gluonic degrees 
of freedom. The important information that is needed to gain better understanding is the extraction of the relevant degrees of 
freedom of the strong interaction in the non-perturbative regime. A much better understanding of the light meson spectrum will be 
very helpful to gain this information and to guide the lattice. 

An important advance will be if one could identify the $s\overline{s}$ states in the light meson sector. There are only a handful
of $s\overline{s}$ states that are unambiguously known. However, they are needed in order to study complete nonets, and to discriminate $q\overline{q}$
against non-$q\overline{q}$ resonances. CLEO-c will attempt to collect in the order of one billion J/$\psi$ decays. This data set will contain some
60 million radiative J/$\psi$ decays. This will be the first data set with large statistics well suited to do meson spectroscopy
in the region of 1.6 to 2.6 GeV. It should enable us to identify most of the $s\overline{s}$ states, since the initial state for radiative J/$\psi$ decays is 
well defined and the CLEO detector will measure all final states simultaneously with 93\% coverage in solid angle.  Another way
of identifying $s\overline{s}$ states is via $\psi$ and $\psi' \rightarrow$VF flavor tagging. Lets say the vector state V might be
reconstructed as $\omega$, $\phi$, or $\rho$ state. The particle F then will be dominantly a state of the same flavor as V 
due to the suppression of hair-pin diagrams.

It has been said many times that identifying the $q\overline{q}$ states is required first in order to find left over states.
Further analysis is needed to determine if they are multiquark states, meson-antimeson molecules, hybrids (qqg) or glueballs.
In practice however, these states can mix with the $q\overline{q}$ state. Current data seem to indicate that we have such a case 
in the scalar sector with the f$_0$(1400), f$_0$(1500) and f$_0$(1710), which are thought to be the mixtures of two $q\overline{q}$ 
states and the glueball, although there are a fair number of other ideas on how to 
identify the scalars. Mixing to such a large degree does not have to be the typical case, and the idea of identifying
the $q\overline{q}$ nonets should usually be possible.

The scalar sector requires special attention. The CLEO-c data set will be the only data set on the horizon that measures all 
three resonances simultaneously with high statistics and a well defined initial state. Furthermore, the scalars should be observed
in $\Upsilon$(1S) decays, with lower statistics but well defined initial states. However the data set that will best reveal the 
nature of the scalar resonances is the 2-photon data collected in the 25fb$^{-1}$ of CLEOII and CLEOIII data, since the coupling 
to glue is suppressed. Therefore the two photon partial width of the scalar states should give us a handle on their glue content.

One also expects hybrid states containing 2 quarks and a gluon as constituents. These are particularly interesting since there 
are hybrid states with quantum numbers not realized by $q\overline{q}$ states. The identification of a spectrum of hybrid states
is very challenging, but has the reward to make unique measurements to test lattice QCD predictions. CLEO-c has  limited access to 
hybrid production from $\chi_{c1}$ decays, enabling us to measure C-odd states as the 1$^{-+}$ exotic hybrid. Since the radiative 
decay fraction of the $\psi'\rightarrow\chi_{c1}$ is about 9\%, a sizeable number of $\chi_{c1}$ decays should be recorded.

\section{A Multitude of other Exciting Measurements}

There are a large number of other measurements possible, which are part of the very diverse CLEO-c program. There are just 
some highlights mentioned here. In $\tau$-physics we expect an improvement in the accuracy of our knowledge of the $\tau$-mass by 
a factor of three and of the Michel Parameter $\eta$ by a factor of four. We expect to detect many rare decays of the D$^0$ and D$^+$ meson
or set limits on the order of a few times 10$^{-6}$. CLEO has the unique ability to perform an R-scan between 3-7 GeV, which would
require about 150 pb$^{-1}$ of data and could be acquired within one week. This is of importance because it is expected that the 
3-7 GeV region will otherwise become the region contributing dominantly to uncertainties in the determination of 
$\alpha$(M$_Z$) and the hadronic contribution to (g-2)$_\mu$.

\section{Conclusions}
The CLEO-c program has many exciting aspects. It is made unique, however, by the fact that it provides the data needed by an emerging
lattice community to validate their new results to a level of a few percent. This combined effort should lead to a situation that
many non-perturbative QCD calculations with well determined errors will be available in flavor physics and wherever they are needed
to explore beyond the Standard Model.


\begin{thebibliography}{citcleoc}

\bibitem{citcleoc}
CLEO collaboration, {\it CLNS 01/1742 at www.lns.cornell.edu/public/CLNS/2001/CLEO.html}, 
CLEO-c and CESR-c: A New Frontier of Weak and Strong Interactions.
\bibitem{citdr}
D. Peterson et al., Proceedings of th 8th Vienna Wire Chamber Conference, 19-23 February, Vienna, Austria, 
to be published in {\it Nucl. Instrum. Methods A}.
\bibitem{citsil}
E. von Toerne et al., {\it hep-ex/0103037}, submitted to {\it Nucl. Instrum.  Methods A}.
\bibitem{citcalo}
T. Hill, {\it Nucl. Instrum. Methods A} {\bf418}, 32 (1998).
\bibitem{citrich} M. Artuso et al. {\it Nucl. Instrum. Meth. A} {\bf 419}, 577 (1998).
\bibitem{citdiscoveretab}
S. Godfrey and J. L. Rosner, EFI-01-10, April 2001, {\it hep-ph/0104253}, submitted to {\it Physical Review D} (Brief Reports).
\bibitem{cithb}
Y.-P. Kuang and T.-M. Yan, {\it Phys. Rev. D} {\bf24}, 2874 (1981); M. B. Voloshin, {\it Yad. Fiz.} {\bf43}, 1571 (1986).
\bibitem{citlat1} 
G. P. Lepage and P. B. Mackenzie, {\it Phys. Rev. D} {\bf48}, 2250 (1993).
\bibitem{citlat2}
See for example: M. Alford et al., {\it Phys. Lett. B} {\bf361}, 87 (1995); M. Luscher et al, {\it Nucl. Phys. B} {\bf478}, 365 (1996).
\bibitem{cithock}
A. Hocker et al., {\it hep-ph/0104062}.
\end{thebibliography}
\end{document}